# ELABORATION OF A NEW TOOL FOR WEATHER DATA SEQUENCES GENERATION

Laetitia Adelard, Mara Thierry, Harry Boyer, Jean Claude Gatina
Laboratoire de Génie Industriel, Université de La Réunion, 15, Avenue René Cassin.
97715 Saint Denis Cedex 9 Reunion Island, France. Tel. 262 96 28 96
Email: adelard@univ-reunion.fr

## ABSTRACT

This paper deals about the presentation of a new software RUNEOLE used to provide weather data in buildings physics. RUNEOLE associates three modules leading to the description, the modelling and the generation of weather data. The first module is dedicated to the description of each climatic variable included in the database. Graphic representation is possible (with histograms for example). Mathematical tools used to compare statistical distributions, determine daily characteristic evolutions, find typical days, and the correlations between the different climatic variables have been elaborated in the second module. Artificial weather datafiles adapted to different simulation codes are available at the issue of the third module.

This tool can then be used in HVAC system evaluation, or in the study of thermal comfort. The studied buildings can then be tested under different thermal, aeraulic, and radiative solicitations, leading to a best understanding of their behaviour for example in humid climates.

## INTRODUCTION

Meteorological data are used in renewable energy management studies for sizing and optimisation of systems (solar heaters, wind generators, so on…). More or less elaborated weather data are needed for buildings simulations. They are used for the evaluation of the thermal comfort, and energy consumption deriving from human discomfort situations (it includes also systems sizing). Thus, a tool providing climatic informations must be able to supply several types of meteorological data :

- "Bin data" usable for systems sizing [Erbs, 83].
- New variables (sky temperature, wet bulb temperature sometimes not available in initial database.
- Representative days or sequences obtained from the existing database.
- Artificial data obtained by using mathematical models, when no database is available.

Such a tool must also propose to the user correlations functions, more or less evaluated models, adapted to the climatic conditions on the studied site. By this way the software can be used in diversified climates.

Generally, the use of meteorological database have been made according to two complementary approaches [Hui and Cheung, 97]:

☞ The first approach is relative to the system sizing (mainly HVAC systems) and uses particularly extreme climatic sequences. These data can then be provided by typical days, by the software elaborated by Van Paassen, or by RUNEOLE. Others techniques exist, for example, the Ashrae organisation method which allows the definition of extreme days in function of their frequency in the year [Ashrae, 93].

☞ The second approach integrates average long term climatic data for the average consumption estimation. Data generally used in this cases are reference years [Chow, 97], or artificial data generated by a software such WETHRGEN elaborated by Degelman [Degelman, 76].

The main objectives and relationships between the two types of database are represented at figure1.

Figure 1 : Relations between the use of mean and extreme data

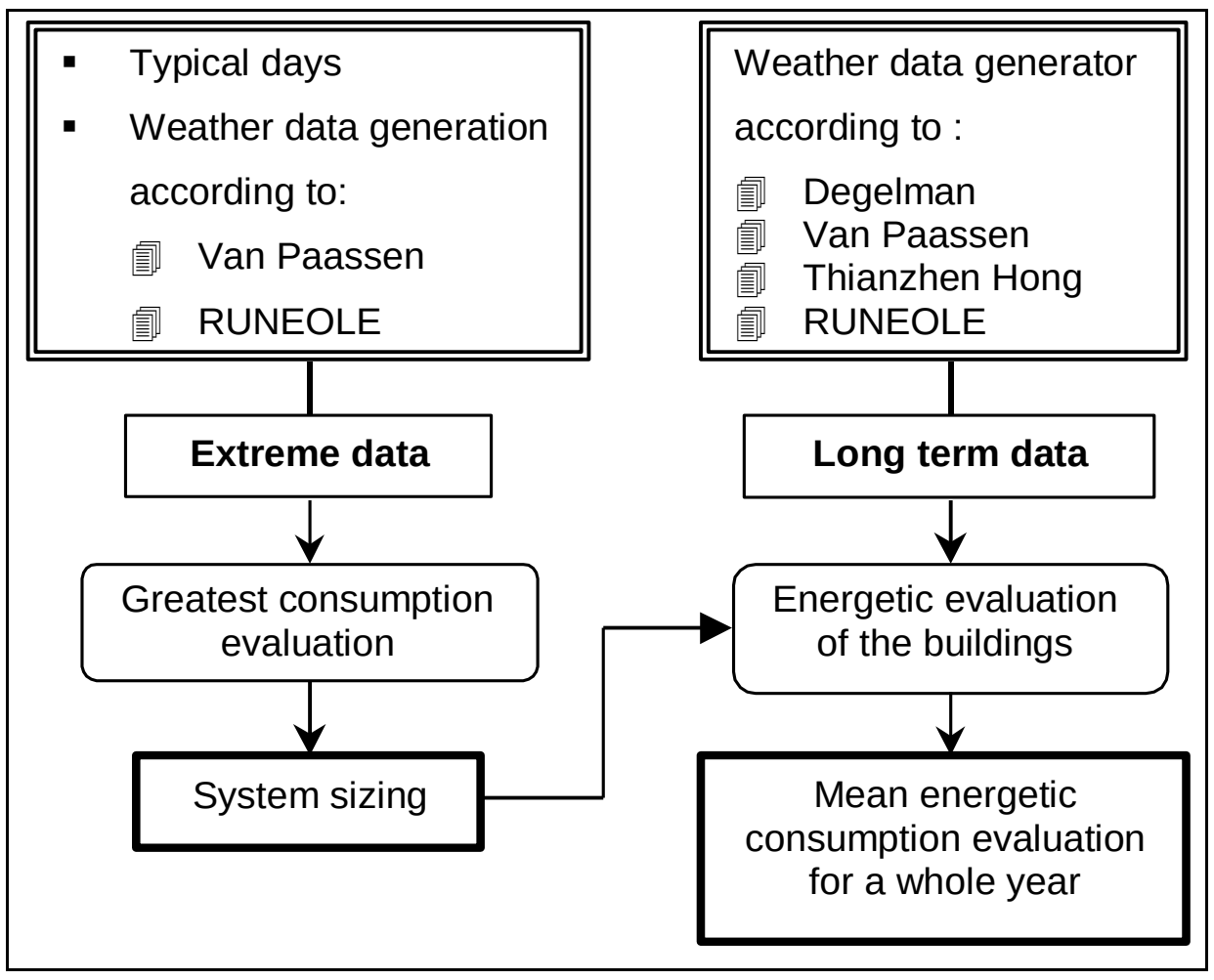

Studies made to compare results of calculations undertaken with typical years, average data, and data measured for the site on fixed duration (a season, a month) [Haberl, 95] confirm this complementarity. For the study of HVAC systems, the use of average data or typical year allow a best estimation of the sensitive power. Measured data allow a best latent power estimation.

Our research aims at the definition of a methodology for the description, modelisation, and generation of specific weather sequences. The software RUNEOLE follows these methodology. The creation of a tool adapted to needs of the searchers and the engineers must be easy to use, interactive. It must be rapidly interfaced with the different computer softwares used by the users. Furthermore, the integration of an expert process by means of a qualitative database with potential for additional facilities can constitute a precious assistance for the user. It allows to guide him in his utilisation of the software, in the choice of mathematics models to use. The proposed tool includes necessary steps for the adaptation of such a software to all types of climates.

The methodological part relative to a detailed analysis of weather sequences leads to a classification of existing climatic situations that can occur on the site. These representative sequences have been used to simulate buildings and HVAC systems thanks to a simulation code, CODYRUN. These methodology for representative weather data determination has been exposed in a previous paper [Adelard, 97]. We will expose here more precisely the modules of modelisation and meteorological data generation of RUNEOLE. We will use the properties of the software MATLAB to build the modular structure of the software, interactive windows, and libraries used for the multimodel approach.

## WEATHER DATA MODELISATION

The modelisation of the climate variables needs an elaborate statistical analysis (figure 2) of the different climatic parameters to go deeply into some aspects of a statistical basis analysis. The modelisation module can be reached at the beginning of the software, or via the basis statistical analysis or data generation modules.

The determination of distribution laws for the climatic variables for hourly or daily data are useful for linear stochastics models elaboration. That is why a library of laws adapted to climatic parameters for different types of climates has been integrated.

Non linear stochastic models with external variables and also simple correlations functions has been programmed in this module to gather all relative tools used for modelisation in the same environment. The software uses automatically this module if a model chosen by the user does not exist in the appointed library to elaborate it.

Figure 2 : Runeole windows of modelisation (Complex statistical analysis).

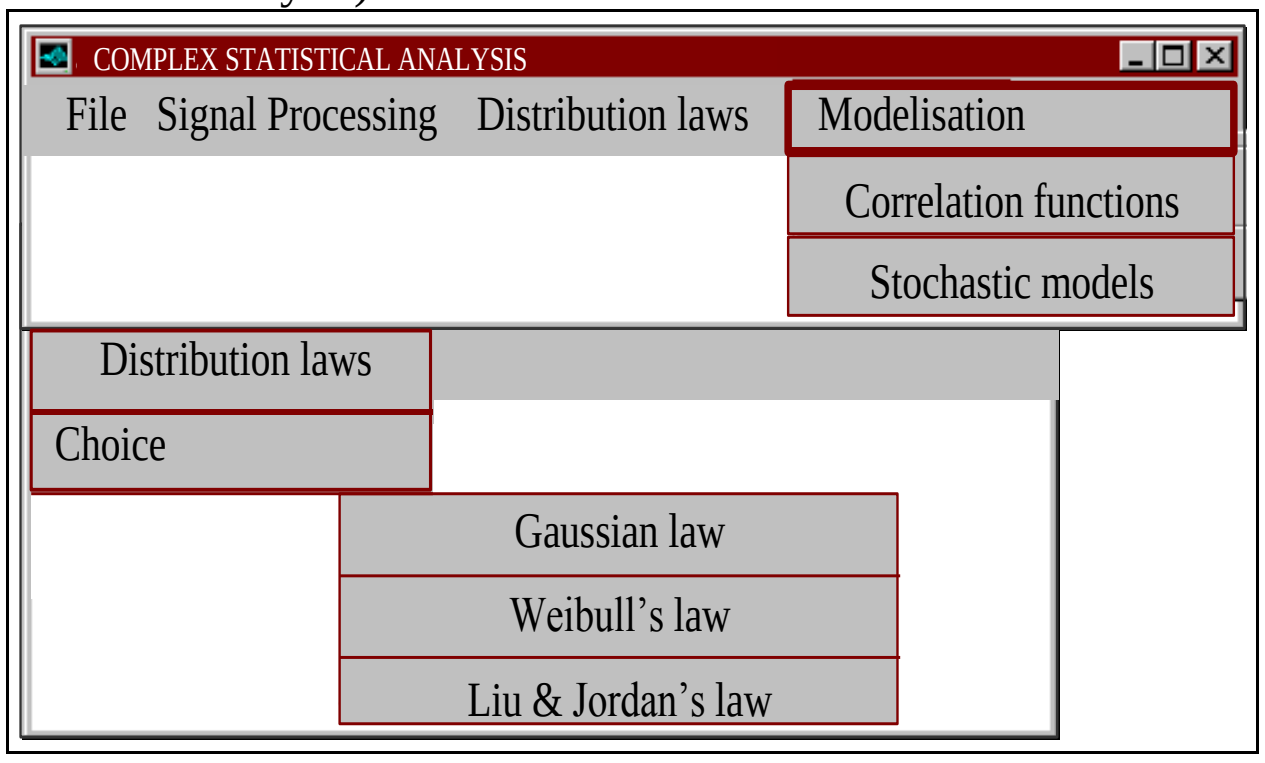

Now we are going to expose some of the results obtained for each steps of modelisation by RUNEOLE.

### Statistical distribution laws

Many publications exposed methods for elaboration of statistical distribution law for wind speed. In most of these publications [Sbai 94], [Blanchard, 84], the Weibull statistical distribution is used to reproduce distribution of wind speed frequencies. The probability function is written:

$$f(v)=\frac{k}{c}\left(\frac{v}{c}\right)^{k-1} e^{-\left(\frac{v}{c}\right)^{k}}$$

with k is parameter form, v is the hourly or daily wind speed and c is a scale parameter .

Figure 3: Graphic window of distribution law results

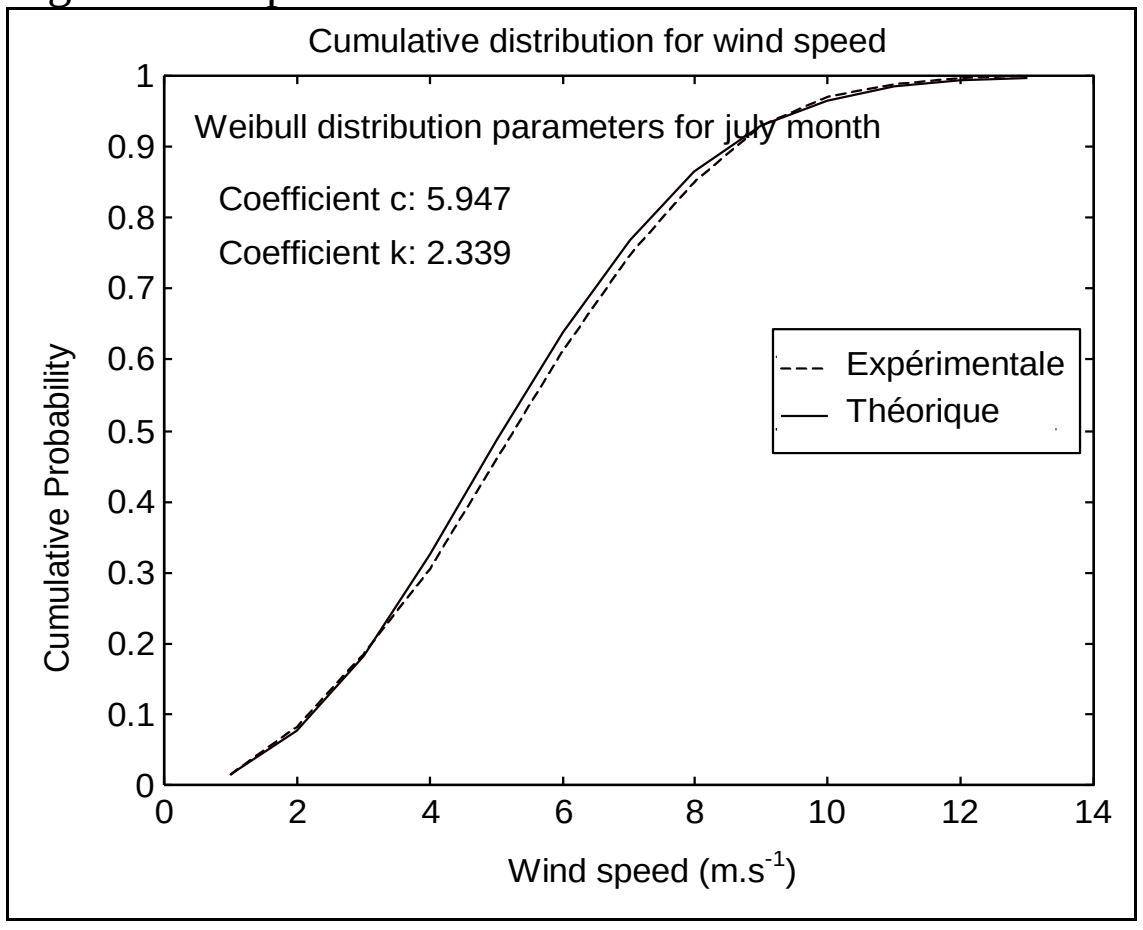

The research of distribution laws in RUNEOLE is made by choosing at first the variable to study, the period, and the theoretical law to use (figure 2). In a second time, RUNEOLE presents the determined coefficients and results in graphic window (figure 3).

Data of radiation are approached by the study of the clearness index, facilitating thus the adaptation to all type of climates. The Liu and Jordan distribution law is usually used, but Saunier [Saunier, 87] demonstrated that this type of relationship does not fit for tropical regions and gives the following formulation:

$$P(x, x_{moy}) = C_1 (x - x_{moy}) e^{\gamma_1 x} \quad \text{with} \quad x = \frac{K_t}{K_{tmax}}$$

Values of $x_{moy}$ and $C_1$ are determined by the next relationships:

$$x_{moy} = \frac{(\gamma_1^2 - 4\gamma_1 + 6)e^{\gamma_1} - 2\gamma_1 - 6}{\gamma_1\left[(\gamma_1 - 2)e^{\gamma_1} + \gamma_1 + 2\right]} \quad \text{with}$$

$$C_1 = \frac{\gamma_1^3}{(\gamma_1 - 2)e^{\gamma_1} + \gamma_1 + 2}$$

$K_{tmoy}$ , $K_{tmax}$ is respectively the daily, or hourly mean and maxi clearness index. Values of $K_{tmoy}$ and $K_{tmax}$ are provided by the user. The coefficient $\gamma_1$ has to be determinate with an iterative process. We have used this law for our site of Gillot, in humid tropical climate (figure 4). Nevertheless, the law of Liu and Jordan [Liu, 60] has also been integrated in the software.

Figure 4: Saunier distribution law applied on daily data of the month of August. Gillot, years 93-96.

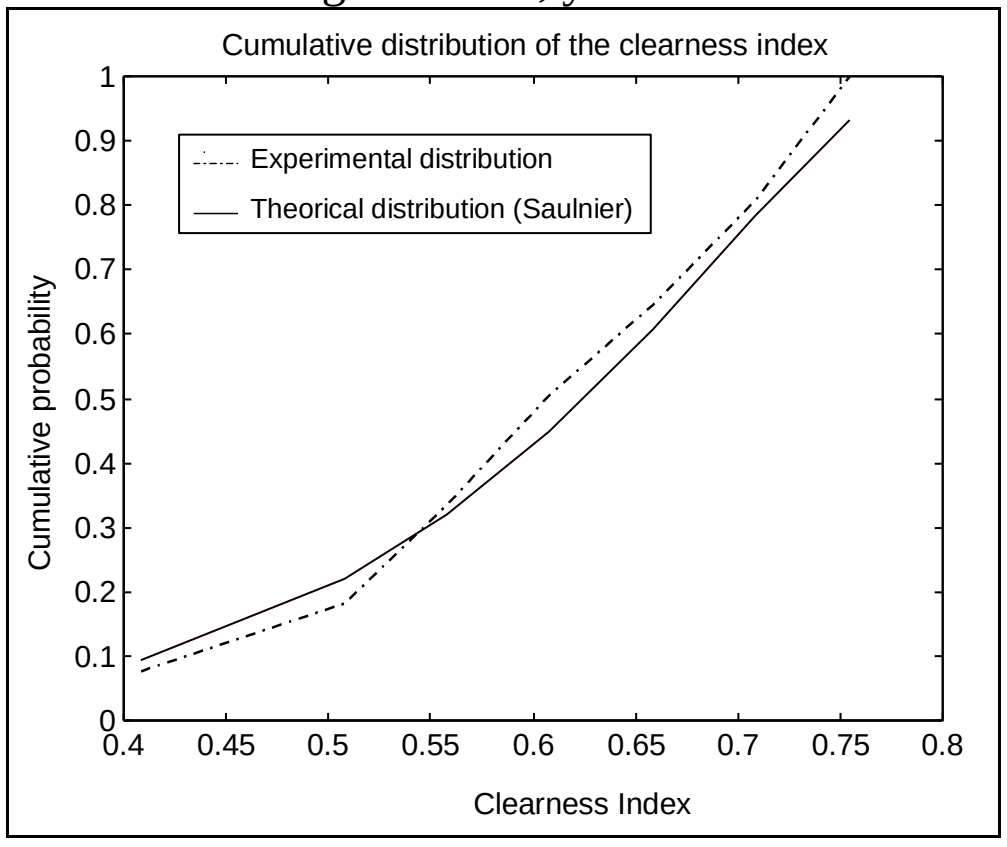

The gaussian law is used for variables such that the temperature and relative humidity.

## Correlations Functions

The correlation functions research part allows to determine these functions for particular climatic sequences. The different functions found in bibliography and relative to solar data have been integrated (table 1). The period, climatic conditions, the variable to predict, the predictors and the type of correlation function has to be chosen, using the window presented in the figure 5. In the same optic that for laws of distributions, the software presents an independent exit window, immediately usable in reports.

Table 1: Correlation functions integrated in RUNEOLE.

| ***Type*** | ***Désignation (Authors)*** |
|---|---|
| Insolation – **Clearness index** | Angström Black |
| | Hay |
| Clearness Index – **insolation** | Angström Black |
| | Hay inverse |
| Global radiation - Insolation - **Diffuse** | Klein |
| | Page |
| Insolation – Global radiation - **Diffuse** | Gopinathan 1 |
| Insolation – **diffuse radiation** | Icqbal |
| Clearness index - Global radiation- | Erbs |
| Clearness index - beam radiation – | Castagnoli |
| Clearness index – **Nebulosity** | Barr |
| Insolation – **Nebulosity** | Rangarajan |
| Global radiation- Insolation - Clearness index -**Diffuse radiation** | Gopinathan2 |
| Wind speed – Global radiation - Diffuse radiation – Temperature | Adelard |
| Insolation – Global radiation- Clearness index - **Diffuse radiation** | Soler |

Figure 5: Window of climatic condition choice for the correlation function research.

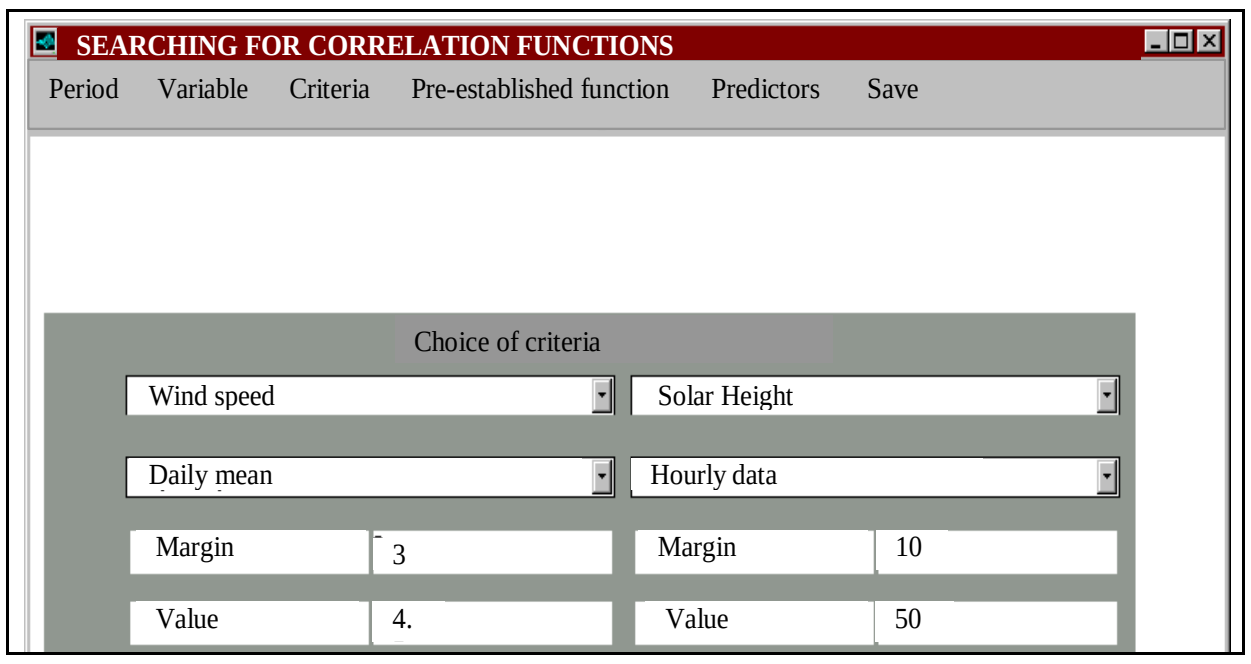

Figure 6 : Error calculated for the function of correlation for an azimuth, and a fixed wind speed.

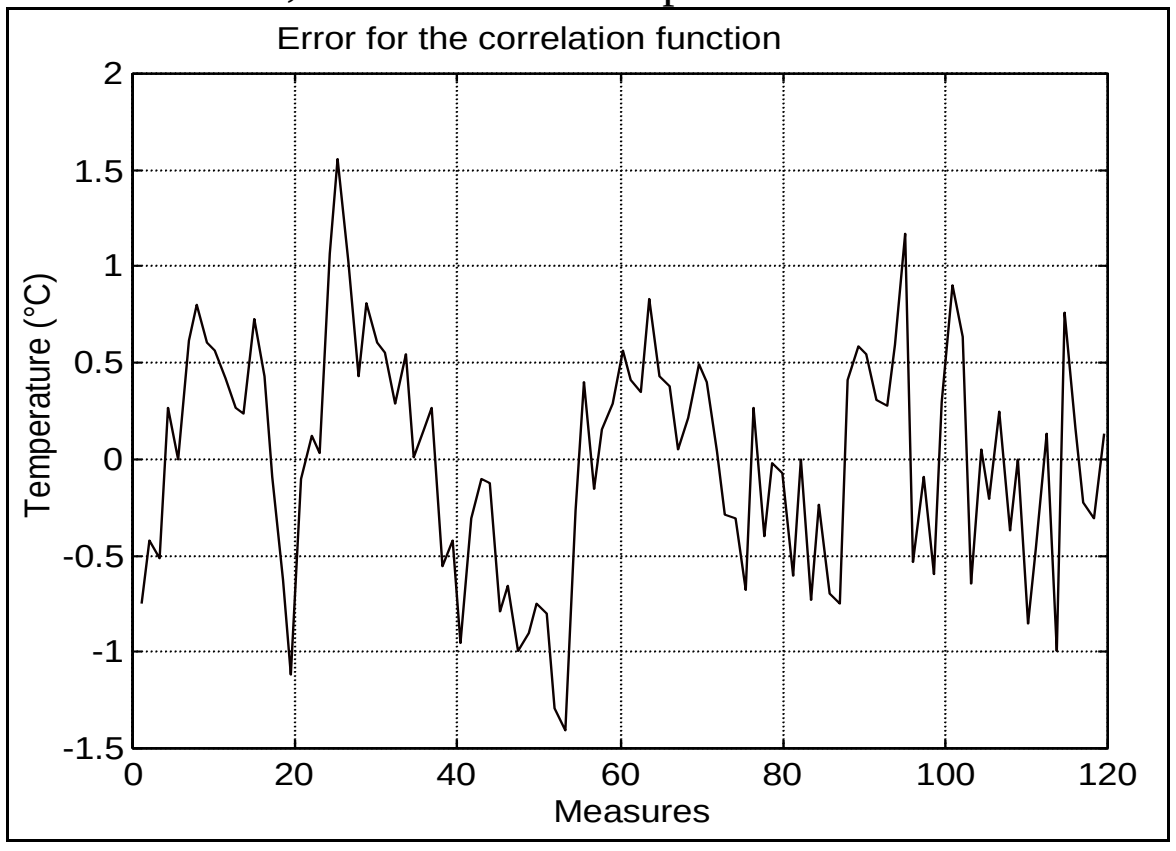

It has been possible to determine polynomial correlation functions linking the temperature to the other climatic parameters in established wind speed conditions, and for different height of sun (these criteria is chosen before searching the correlation function) (figure 6).

Nevertheless, it is appeared that these correlation functions give satisfying results for months of humid season, for the site of Gillot (Reunion Island), but not for climatic conditions of the dry season. The determination of reliable model representing interactions between the temperature, relative humidity and the other variables such that the wind speed and the solar radiation includes the use of more elaborated techniques such that neural networks for the creation of non linear models.

On the other hand, if the user wishes to generate data by fixing wind and radiation conditions, it is necessary to generate these variables independently, leading us also to the creation of linear autoregressive models. These models will be described in the next part.

## Determination of autoregressive models

Stochastic models, generally called ARMA (AutoRegressive Moving Average models), allow to write the variable X at the instant n according to its state at the instant n - τ with the relationship:

$$X(n) = \sum_{\tau=1}^{p} \phi_\tau . X(n-\tau) + w(n) - \sum_{\tau=1}^{q} \theta_\tau . w(n-\tau)$$

- n et n-τ are number of days or hours.
- X(n) is the value of X for the day or the hour n
- w(n) et w(n-τ) are random values of a variable of Gauss of average 0.
- $\phi_\tau$ represent autoregressive parameters. They are determined by studying the autocorrelation of the variable. $\theta_\tau$ represent the moving average parameters.

The determination of these models lays in the determination of coefficients and p and q associated orders as well. A series of steps allows us to create the model:

1. Determination of the nature of the model: we have to choose between the elaboration of a purely autoregressive models, a purely moving average, or mixed model. This part is made taking into account the partial and initial autocorrelation of the studied variable [Box, 76] (figure 7). The partial autocorrelation are determined by relationships of Durbin [Box, 76].
2. Determination of orders p and q. This determination is made using the Bartlet test for the initial autocorrelation and the Quenouille test for the partial autocorrelation.

These two tests allow to study the nullity of these two variables, and to choose orders p and q according to their variations.

3. Estimation of coefficients of the model.
4. Verification of the model by test of the autocorrelation of the residue (the autocorrelation of the residue must be null).

Figure 7: Partial and initial autocorrelations for the speed of the wind (Month of August).

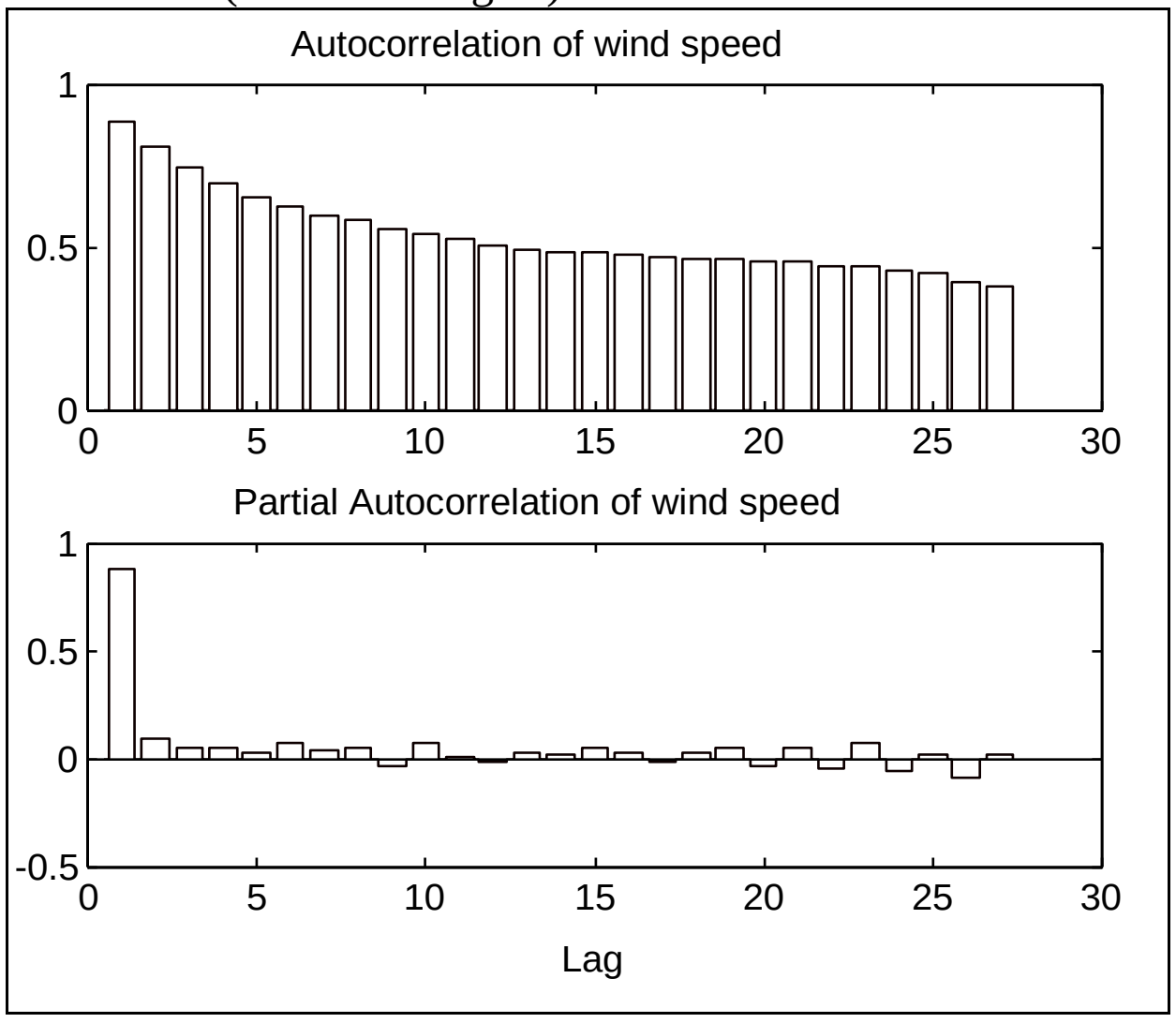

This kind of model has been tested for the generation of wind speed giving some good results (figure 8).

Figure 8 : Result window for the stochastic modelisation (autoregressive model elaborated for the wind, hourly data, site of Gillot).

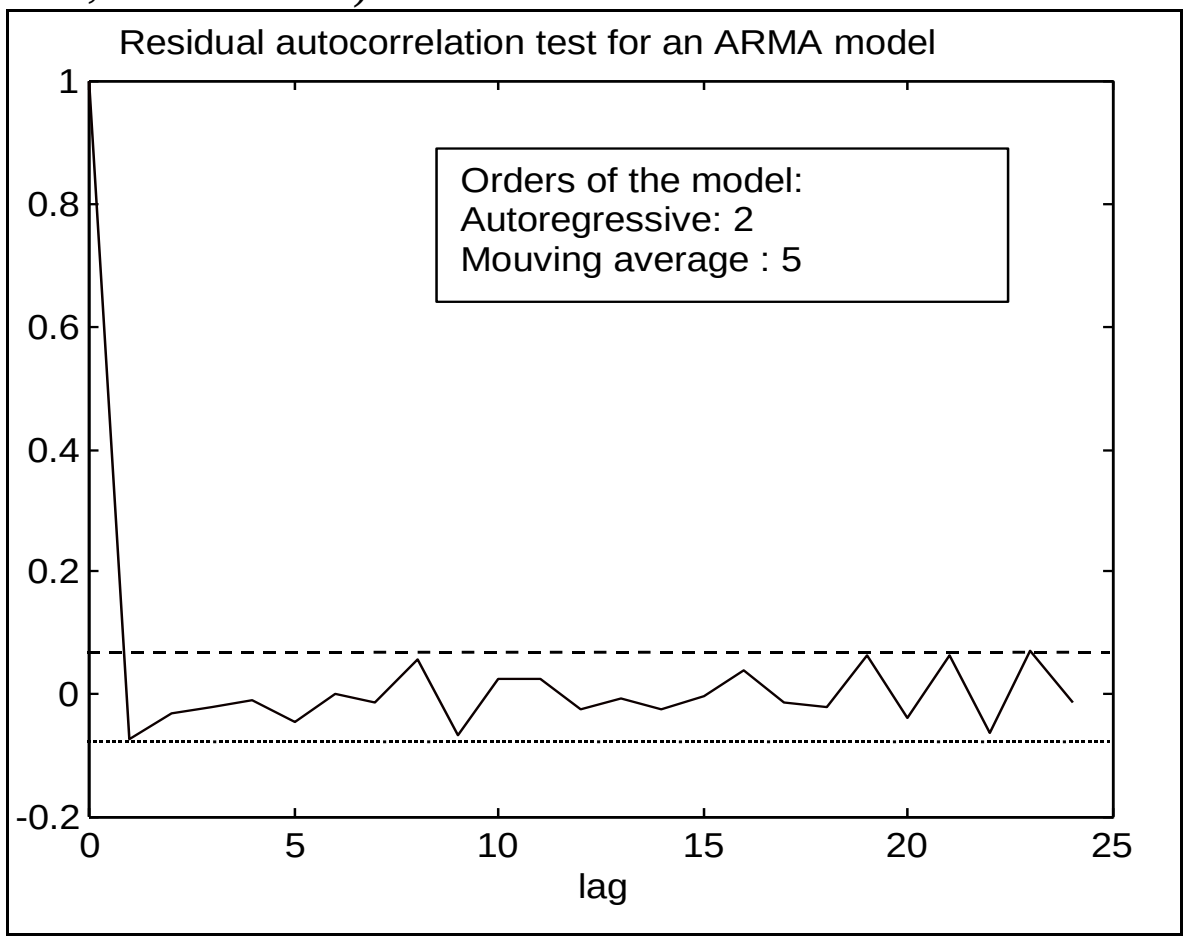

For the taking into account of interactions between the different climatic variables, we have used the technique of neural network which have been used in various fields (recognition of forms, process control [Grondin Perez, 94]; [Nefati, 93], meteorology [Buffa, 98]; [Marzban, 96])…

It can be used in the case of noisy data, and allows the identification of system on a wide functioning range.

The neural network is constituted of linear or non linear functions associated, forming thus a constituted system of several layers (generally two layers, the first is the input layer and the second one integrates the output function). To each entry of these functions is associated a pondered weight determined from a learning phase. Create a model for climatic variable from a neural network includes several phases.
1. Choice of entry variables (that will be for us hourly data of climatic parameters).
2. Choice of the function of activation and the architecture of the network.
3. Choice of the algorithm allowing to determine the different connection weight. The algorithm of Levenberg Marquardt that we will use is a method derived of the approach Gauss-Newton used for the optimisation of the quadratic error (between the measured and the calculated data).

From solicitation variable determined in the signal processing tools, we have determined entries of the network. The number of necessary hidden neurones for the model is found after successive tests. The evaluation of the model is made according to the quadratic mean error (eqm) calculated between the measured exit s(k) and the calculated exit there(k), during the learning phase. Systems have been used here for the generation of humidity data according to the temperature, and data of temperature in function of the global radiation.

Figure 9 : Error of the model obtained by neural network.

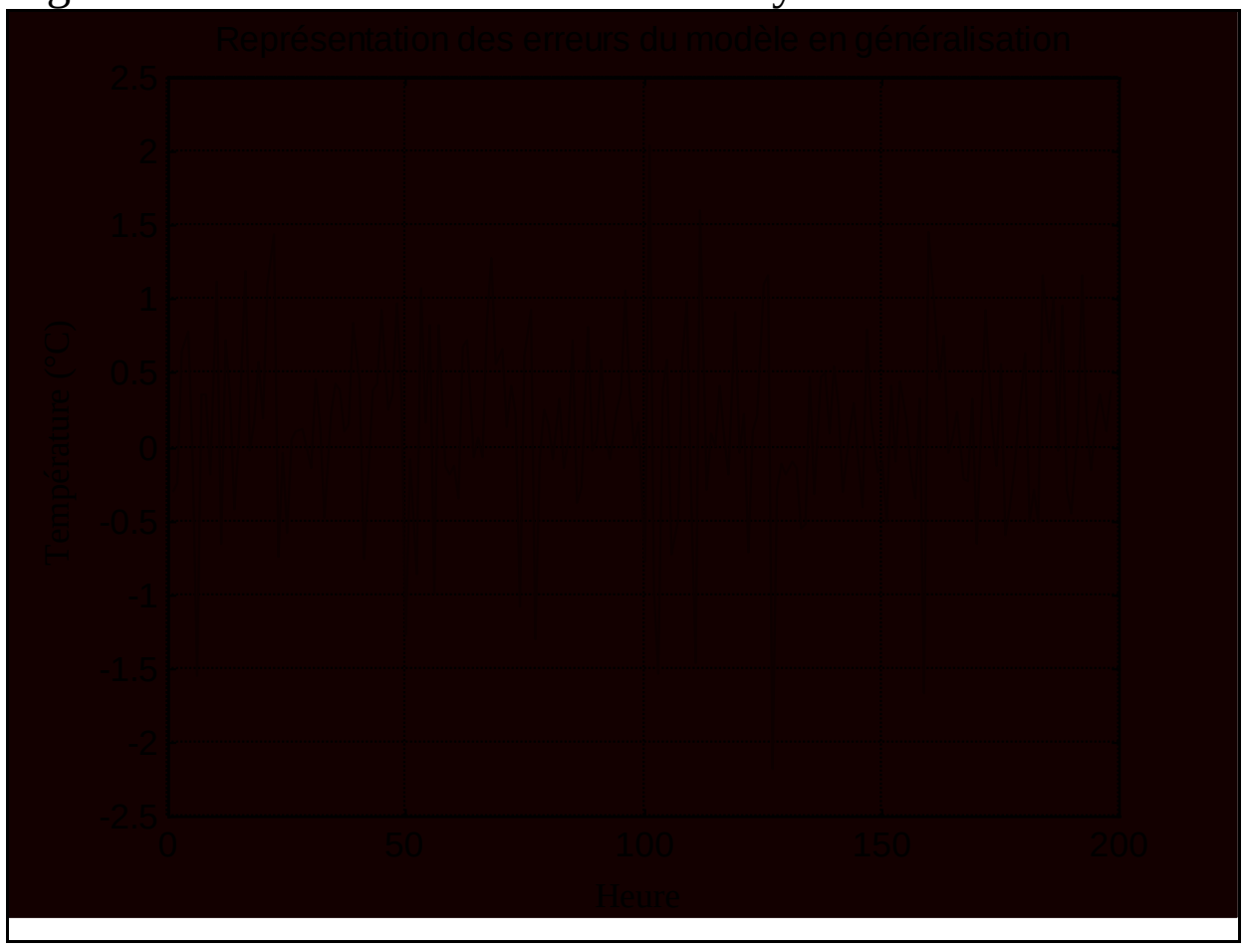

The predictors chosen for the temperature are the diffuse, global solar radiation, and the wind speed. The retained system is a network with 3 hidden neurones. We have used the algorithm of Levenberg Marquardt [Noorgaard, 96] for the determination of weights of connected them. Results found by this method are close to measured data. The average observed error is about 2% in positive value and 3.5% in negative value (figure 9). An analysis of the sensitivity of the error considering the variables used as input will be made to improve the model.

## CLIMATIC SEQUENCE GENERATION

The bibliographic study allowed to identify five main algorithms for climatic sequence generation. They are generally composed of a statistical part allowing the generation of the purely random aspect of the climate from distribution laws and determined stochastic models, and of a part based on empirical relationships or on autoregressive models with external variables to generate the other climatic parameters. The order of generation of data is beforehand fixed by the structure of the algorithm of model association. Van Paassen [Van Paassen, 79] generates thus the global radiation and in a second time the other climatic parameters taking into account a random part and a part linked to the global radiation. The same concept is found with Degelman. The intercorrelation of parameters is little processed in its works [Degelman, 76; 91]. Its step has been partly modified by Knight [Sel, 90]. Thianzhen Hong [Thianzhen Hong, 95], generated data concern solely the temperature and the global solar radiation.

The climatic sequence generation lays for the essential on the complex statistical and quantitative analysis that have been made in the modelisation part. It uses notably anterior models stocked in library. If the user wishes to have a particular climatic sequence, duration, or different type of climatic sequences which does not exist in available meteorological files at its disposition, RUNEOLE has then to propose variables to be generated and criteria for climatic conditions choice for the generation of the new sequence (figure 10). From now, the software chooses models to use and proposes a choice to the user if necessary (see algorithm at figure 11).

Figure 10 : Generation of artificial climatic sequences

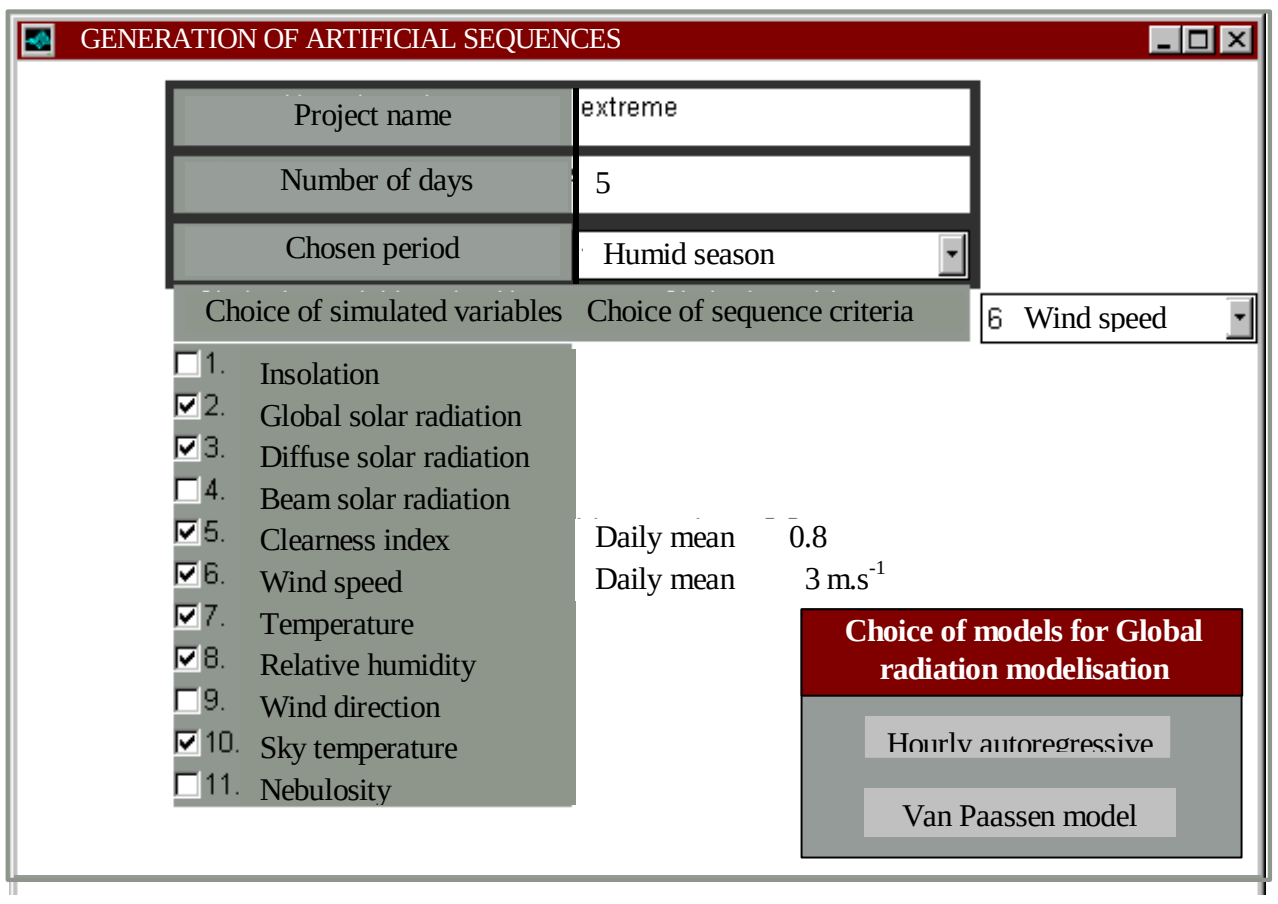

The software includes a file that contains designations of models being able to be elaborate or used, variables used as predictors, and programs used for the elaboration of these models. It exploits this program to know what are models it can use, or that it can propose to the user. In the case of an empty library, the software goes automatically to the complex statistical analysis module, which allows to create correlation functions, or more complex models. The choice of models to be used is made with the help of a choice window (figure 10).

Figure 11 : Algorithm of models and generation association of sequences.

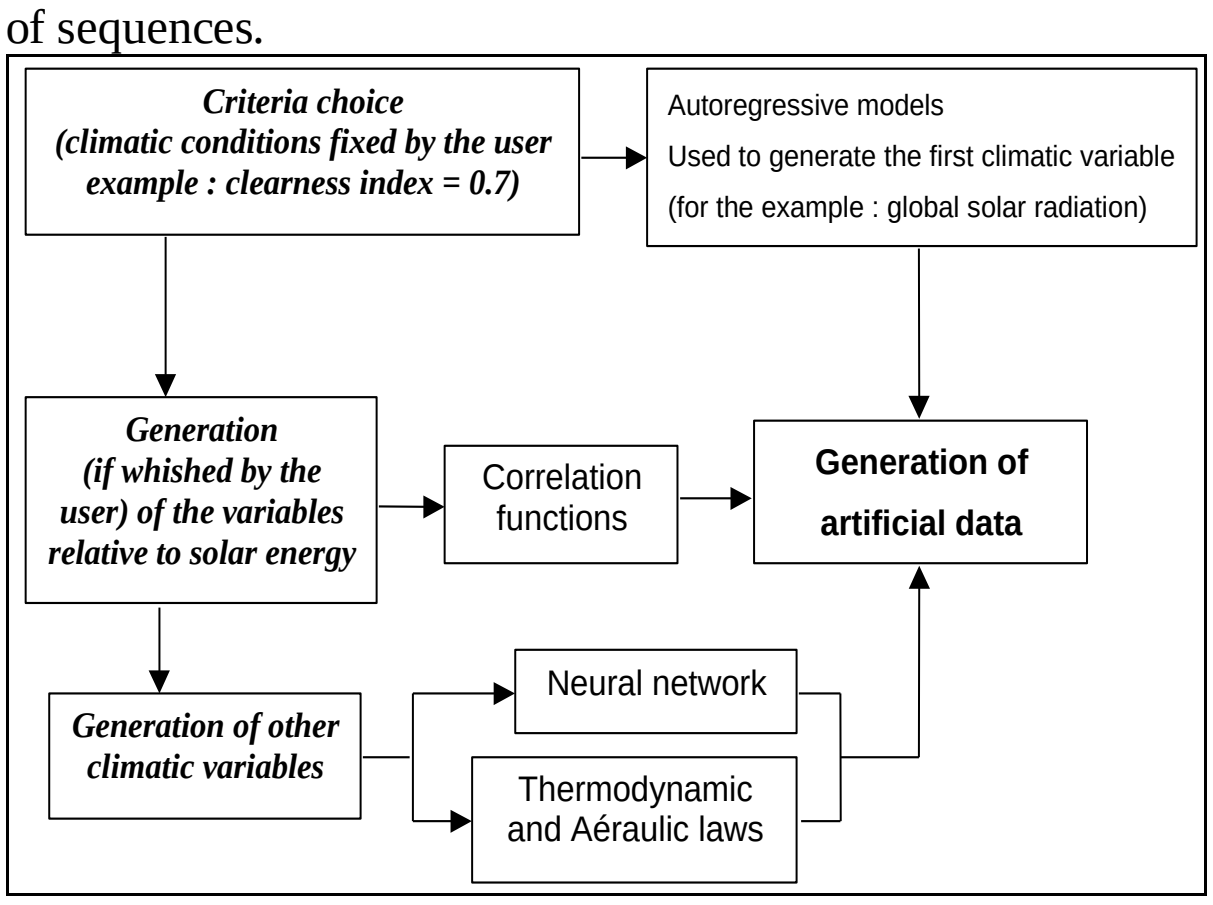

Climatic data are therefore generated in function of models used. It is then possible to export the generated data. The file exportation for use with other software is governed by a program.

Figure 12 : Windows for the reading of generated data.

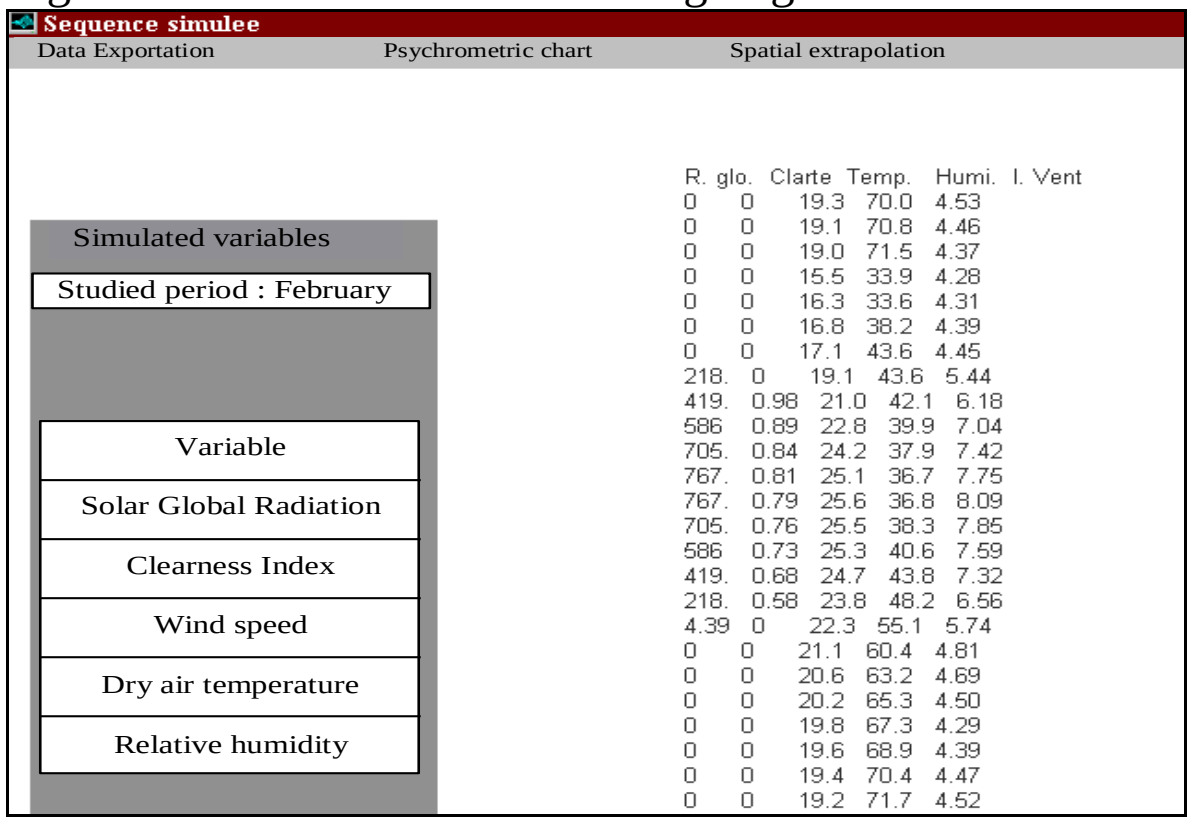

It is necessary to remind that in our case, generated sequences have to correspond to criteria fixed by the user, i.e. to conditions fixed for some climatic variables. These variables will be therefore generated from autoregressive models. Variables relative to the solar radiation such that the insolation, the diffuse, direct solar radiation and the nebulosity can be obtained thanks to established correlation functions.

Thermodynamic variables (temperature and humidity) will be generated using "black box" models. The established algorithm insures thus the coherence between the different climatic variables generated. The algorithm of sequences generation and assembly of models is presented in the figure 11 and the climatic sequence generated in the figure 12.

## VALIDATION

The objective now is to validate such a software. Some indirect or direct step process have been elaborated. The direct step lays on statistical models and distribution laws verification. The validation has then to be made to two levels :

### Elaborate model validation

Elaborate models are validated by comparison of measured and simulated data. This type of validation lays on the three main statistical tests that are tests of correlations, tests of law comparisons of distributions, and tests of autoregressive models. We present in the table 2 the main tests used in these optics.

Table 2 : Tests applied on the validation of models according to the various authors

| Models to validate | Statistical Test |
|---|---|
| Correlations functions | Fischer Test on the coefficient of correlation<br>Tests of coefficients of the regression |
| Statistic distribution laws | Chi - 2 test for the comparison of the theoretical distribution, and the real distribution |
| Stochastic model | Test of autocorrelation |
| Non linear Models | Evaluation by the quadratic average error |

### Validation of generated data

At this level, the simulated data coherence is a great imperative. The test used for the comparison of the two simulated and measured statistical distributions is the two sided Kolmogorov-Smirnov test.

The validation at daily and hourly level lays on the extreme value verification and frequencies of occurrence of some climatic parameters such that dry air temperature. The hourly data validation of temperature has been undertaken by verifying the combined evolution of dry and wet bulb temperatures. Extreme insolation values have also been tested. At the monthly level, the validation lays on a simple mean and standard deviation comparison for simulated and measured data.

The indirect method consists in exploitations of the building simulations to test the generated climatic data coherency. The generated data and the measured data are used in simulation to study the behaviour of for example an HVAC system. The similarity between internal temperatures of the systems and the ratio of overheats is then tested, always with the Kolmogorov-Smirnov test.

The very deep validation of the RUNEOLE as generator of data, would constitute an important work that could be the subject of ulterior developments. The analysis and characterisation parts are generally not include in a classic software validation based on a comparison of data coming from various sources and the parametric sensitivity analysis. Its validation could have lean on an accumulation of expert knowledge that would confirm the analysis and conclusion pertinence obtained in these parts.

## CONCLUSION

We have exposed in this paper the elaborated software combining the three steps of characterisation, modelisation, and climatic sequence generation. The expert approach associated with this software by means of analysis windows accompanying the user in his work, can constitute a non negligible assistance. Perspectives of this work are multiple, as the creation of a more "universal" software programmed in C language, and especially the programming of the module of physical modelisation. The modular structure of the software allows a constant and rapid evolution of the different module. The multimodel approach lays on the creation of libraries adapted to waits of searchers, automatic representative days or sequences. We have also integrated all interesting statistical tools to drive the necessary mathematical analysis. These approaches more or less classic, more or less elaborated, have been used of always partial manner by several authors. They have been, in major part, integrated in RUNEOLE because we have wished to have a totality of quasi exhaustive tools to characterise and model the climatic data. We have shown in several applications the interest of this step [Adelard, 98].

The next steps of this work will allow to test the software on others meteorological files so as to improve some distribution laws or correlation, and to develop the hybrid modelisation. On the other hand, it is necessary to continue the integration of physical model leading to some spatial extrapolations of microclimatic data, taking into account of the altitude, the dominant wind direction or the particular characteristics of the relief.

At the level of applications in building physics, more elaborate analysis of sensitivity of parameters between the output variables of the different energy system, and climatic solicitations have to be undertaken. A new simulation approach is now possible here. We can indeed observe the reply of a building for different climatic solicitations on a building, in the same spirit that simulations driven with variable structural parameters.

## BIBLIOGRAPHY

[Adelard, 1997] L. Adelard, F. Garde, F. Pignolet-Tardan, H. Boyer, J C Gatina, Weather sequences for predicting HVAC system behaviour in residential units located in tropical climates, *IBPSA Congress Proceedings, Prague, 1997*

[Adelard, 1998] L. Adelard, Caractérisation de bases de données climatiques, Proposition d'un générateur de climat, Applications en thermique de l'habitat, Thèse de troisième cycle, Sci.

[Ashrae, 1993] ASHRAE, Fundamentals Handbook, *Ashrae Inc., 1791, Tullie Circle, Atlanta, GA 30329*

[Barr, 1996] A. G. Barr, S.M. MCGIN, SI BING CHENG, A comparison of methods to estimate daily global solar irradiation from other climatic variables on the canadian prairies. *Solar Energy, Vol. 56, No. 3, p. 213-224 (1996)*

[Blanchard, 1984] M. Blanchard, et G. Desrochers, Generation of autocorrelated wind speeds for wind energy conversion system studies, *Solar Energy, Vol. 33, N°6, p. 571-579, 1984*

[Box, 1976] G.E.P. Box, G.M. Jenkins, Time series analysis: Forecasting and control, *Revised Edition, Holden Day, 575 p., 1976*

[Buffa, 1998] F. Buffa, I. Porcedu, Temperature Forecast and dome seeing minimization-I. A case study with a neural network model, TNG Technical report N°67

[Chow, 1997]W.K. Chow, S.K. Fong, Typical meteorological year for building energy, *Architectural Science Review, Vol. 40, N°1, p. 11-15*

[Degelman, 1976]L. O. Degelman, A Weather simulation model for Building Analysis, *ASHRAE Trans., Symposium on Weather Data, Seattle, WA,435-447*

[Degelman, 1991] L. O. Degelman, A statistically - based hourly weather data generator for driving energy simulation and equipment design software for buildings, *2nd world congress of technology for improving the energy use, comfort, and economic of building worldwide, International Building Performance Simulation As., Nice, Sophia-Antipolis, 1991*

[Erbs, 1983] D.G. Erbs, S.A. Klein, and W. A. Beckman, Estimation of degree-days, and ambient temperature bin data from average temperatures, *Ashrae Journal 25, N°6, p. 60-65, (1983)* [Gopinathan, 1992], K.K.Gopinathan, Solar Sky estimation techniques, *Solar Energy, Vol.49, N°1, p. 9-11, 1992*

[Grondin Perez, 1994] Brigitte Grondin Perez, Les réseaux de neurones pour la modélisation et la conduite des réacteurs chimiques: simulations et expérimentations., *Thèse Sci.; Université de Bordeaux I, 1994*

[Haberl, 1995] Jeff S. Haberl, Doug J. Bronson, Dennis L. O'Neal, Impact of using measured data vs. TMY weather data in a DOE-2 simulation., *ASHRAE Trans., (1995).*

[Hay, 1979] J.E. Hay, Calculation of monthly mean solar radiation for horizontal and inclined surfaces, *Solar Energy, vol.23, p. 301-307, 1979*

[Hui, 1997] Hui, S C M, K. P. Cheung, Multi-Year (MY) buildings simulation: is it useful and practical?, *IBPSA congress proceedings, p.. 277-284, Prague, 1997*

[Iqball, 1979] M. Iqball , A study of Canadian diffuse and total solar radiation data monthly average daily horizontal radiation, *Solar Energy, vol.22, p. 81-86, 1979*

[Klein, 1977], S.A. Klein, Calculation of monthly average insolation on tilted surfaces, *Solar Energy, vol. 19, p. 325-329, 1979*

[Liu, 1960] B.Y.H. Liu, R.C. Jordan, The interrelationship and characteristic distribution of direct, diffuse, and total radiation., *Solar Energy 4 (3), 1-19 (1960)*

[Marzban, 1996] C. Marzban, G. J. Stumpf, A neural network for tornado prediction based on doppler radar-derives attributes, *Journal of applied meteorology, Vol. 35, 617-626*

[Matlab, 1992] The MathWorks, Inc., Matlab reference guide, *1992, 548p.*

[Nefati, 1993] H. Nefati, J.J. Legendre, C. Michot, Prédiction de la sensibilité au choc des explosifs. Comparaison des approches statistique et neuronale, *Europyro 1993, p. 79-87*

[Noorgaard, 1996] Norgaard M., System identification and control with neural networks, *Thesis, Department of Automation , Technical University of Denmark.*

[Page, 1964], J. K. Page, The estimation of monthly mean values of daily total short-wave radiation on vertical and inclined surfaces from sunshines records for latitudes 40°N-40°S, *Proceedings of the UN Conference on New Sources of Energy, 4, 378 (1964).*

[Saunier, 1987] G.Y. Saunier, T.A. Reddy, S. Kumar, A monthly probability distribution function of daily global irradiation values appropriate for both tropical and temperate locations., *Solar Energy , Vol. 38, No. 3, Pages 169-177, (1987)*

[Sbai, 1994] A. Sbai, A. Moudhi, N. Adouk, et F. Paul, Modélisation de la vitesse du vent et calcul du potentiel éolien du Maroc oriental, *La météorologie, N° 5, 8 éme série, mars 1994*

[SEL, 1990] Solar Energy Laboratory, TRNSYS, a transient simulation program, *Engineering experiment station report 38-13, Septembre 1990*

[Thianzhen Hong, 1995] T. Hong, Y. Jiang, Stochastic Weather Model for Building HVAC Systems, *Building and environment, Vol. 30, N°4, p. 521-532, 1995*

[Van Paassen, 1979] A.H.C. Van Paassen and A.G. Dejong, The Synthetical Reference Outdoor Climate, *Energy and Buildings, Vol (2) Pages 151-161*